\documentclass[aps,prd,reprint,preprintnumbers,showpacs,showkeys,superscriptaddress,nofootinbib,floatfix]{revtex4-1}
\usepackage{amsmath,amssymb,bm,mathrsfs}
\usepackage{times}
\usepackage{epsfig}
\usepackage[colorlinks=true,citecolor=blue,linkcolor=blue]{hyperref}
\usepackage{graphicx}
\usepackage{subfigure}
\usepackage{multirow}
\usepackage[perpage,symbol*]{footmisc}
\usepackage{slashed}
\usepackage{simplewick}

\def\chm{\checkmark}

\begin{document}

\title{Universal Asymptotic Eigenvalue Distribution of Large $N$ Random Matrices\\--- A Direct Diagrammatic Proof of Marchenko-Pastur Law ---}

\author{Xiaochuan Lu}
\email{luxiaochuan123456@berkeley.edu}
\affiliation{Department of Physics, University of California,
  Berkeley, California 94720, USA}
\affiliation{Theoretical Physics Group, Lawrence Berkeley National
  Laboratory, Berkeley, California 94720, USA}

\author{Hitoshi Murayama}
\email{hitoshi@berkeley.edu, hitoshi.murayama@ipmu.jp}
\affiliation{Department of Physics, University of California,
  Berkeley, California 94720, USA}
\affiliation{Theoretical Physics Group, Lawrence Berkeley National
  Laboratory, Berkeley, California 94720, USA}
\affiliation{Kavli Institute for the Physics and Mathematics of the
  Universe (WPI), Todai Institutes for Advanced Study, University of Tokyo,
  Kashiwa 277-8583, Japan}

\begin{abstract}
In random matrix theory, Marchenko-Pastur law states that random matrices with independent and identically distributed entries have a universal asymptotic eigenvalue distribution under large dimension limit, regardless of the choice of entry distribution. This law provides useful insight for physics research, because the large $N$ limit proved to be a very useful tool in various theoretical models. We present an alternative proof of Marchenko-Pastur law using Feynman diagrams, which is more familiar to the physics community. We also show that our direct diagrammatic approach can readily generalize to six types of restricted random matrices, which are not all covered by the original Marchenko-Pastur law.
\end{abstract}
\preprint{UCB-PTH-14/36, IPMU14-0317}
\maketitle

\section{Introduction}

Random matrix theory is widely used in the study of theoretical physics. And the large $N$ limit has been a very useful tool, both at a qualitative and a quantitative level~\cite{coleman1987aspects}. It works as a great approximation, even when the actual dimension of the matrix is not a very large number, such as $N_c=3$ in QCD \cite{'tHooft:1973jz,Manohar:1998xv} and $N_f=3$ in neutrino anarchy~\cite{Hall:1999sn,Haba:2000be,Bai:2012zn,Lu:2014cla}. Therefore, the behavior of random matrices under large $N$ limit can provide insight to many theoretical studies in physics.

Often in this kind of studies, the behavior of eigenvalues are of special interests to us, as they usually represent crucial quantities of the model, such as masses in cases of particle physics. But of course the eigenvalue distribution generically depends on the prior of the random matrices, and there is no privileged choice. However, under large dimension limit, a powerful theorem---Marchenko-Pastur (MP) law--- states that there is a universal asymptotic eigenvalue distribution as long as all the entries are independent and identically distributed (i.i.d), regardless of the choice of the entry distribution~\cite{marchenko1967distribution}. To be more concrete (still a sketch here, see Section~\ref{sec:MP} for the precise statement), let $X$ be a random $M\times N$ matrix with $M\times N$ i.i.d. complex/real entries, then under large $N$ limit, the asymptotic eigenvalue distribution of the matrix $A=\frac{1}{N}XX^\dagger$ is universal and called Marchenko-Pastur distribution. The theorem is named after Ukrainian mathematicians Vladimir Marchenko and Leonid Pastur who proved this result in 1967~\cite{marchenko1967distribution}. After the original paper, a lot of works followed and the theorem is sharpened into a few different versions, each of which has a different set of premises assumed (see \textit{ e.g.}~\cite{gotze2004rate} for a brief summary of the story).

The key point of the MP law is the universality, namely that any i.i.d. entry distribution (within some restrictions, see Section~\ref{sec:MP} for details) will yield the same asymptotic eigenvalue distribution. Although this law got proven almost 50 years ago, one of the new contributions of this paper is to provide an alternative proof of it---a direct diagrammatic method which is more familiar to the particle physics community, so that this universality can be better understood. In addition, the original MP law only covers the case of $X$ being an arbitrary complex/real matrix with $M\times N$ i.i.d. complex/real entries. But in many physics models, cases of restricted $X$ are of interests, such as symmetric, antisymmetric, Hermitian, \textit{etc}. For these types of restricted $X$, is the large $N$ eigenvalue distribution of $A=\frac{1}{N}XX^\dagger$ still MP distribution? With the direct diagrammatic method presented in this paper, one can answer this question easily. As we will show in Section~\ref{sec:Generalizing}, our method generalizes to six types of restricted $X$ with little effort, making it very transparent that the MP universality should hold for all of the following seven cases of $X$:
\begin{eqnarray}
&&\text{(1) Complex arbitrary} \nonumber \\
&&\text{(2) Complex symmetric} \nonumber \\
&&\text{(3) Complex antisymmetric} \nonumber \\
&&\text{(4) Real arbitrary} \nonumber \\
&&\text{(5) Real symmetric} \nonumber \\
&&\text{(6) Real antisymmetric} \nonumber \\
&&\text{(7) Hermitian} \nonumber
\end{eqnarray}
with case (1) and (4) being the original MP law.

\renewcommand\arraystretch{1.4}
\begin{table*}[tb]
\centering
\begin{tabular}{|c|ccccccc|ccccccc|}\hline
               & \multicolumn{7}{|c|}{i.i.d. Gaussian} & \multicolumn{7}{|c|}{i.i.d non-Gaussian} \\
  \hline
  cases of $X$ & (1) & (2) & (3) & (4) & (5) & (6) & (7) & (1) & (2) & (3) & (4) & (5) & (6) & (7) \\
  \hline
  our work in this paper
  & \chm & \chm & \chm & \chm & \chm & \chm & \chm & \chm & \chm & \chm & \chm & \chm & \chm & \chm \\
  MP Law~\cite{marchenko1967distribution}
  & (\chm) & & & (\chm) & & & & (\chm) & & & (\chm) & & & \\
  Wigner's Semicircle Law~\cite{wigner1993characteristic,wigner1958distribution}
  & & & & & (\chm) & & & & & & & (\chm) & & \\
  \cite{PhysRevE.49.2588,PhysRevE.53.1399,Zee:2003mt}
  & & & & & \chm & & \chm & & & & & \chm & & \chm \\
  \cite{Feinberg:1996qq}
  & \chm & & & (\chm) & & & (\chm) & (\chm) & & & (\chm) & & & (\chm) \\
  \hline
\end{tabular}
\caption{Comparison between our work in this paper and some of the closely related works in the literature. For each work, we put a ``\checkmark'' if a case is proved by a diagrammatic method, a ``(\checkmark)'' if the case is proved by a non-diagrammatic method, and leave it blank if the case is not discussed.} \label{tbl:Comparison}
\vspace{-10pt}
\end{table*}
\renewcommand\arraystretch{0}

After finishing this paper, we learned that both asymptotic eigenvalue distribution and diagrammatic methods had been discussed extensively in the literature. And some of the works are closely related to our work in this paper. So it is useful to comment on the relevances and detailed differences. Table~\ref{tbl:Comparison} summarizes a comparison between our work and several closely related literatures, from which one can see the novel aspects of our work and how it serves as a complementary approach to many other works.

First, a very well known result is the Wigner's semicircle law~\cite{wigner1993characteristic,wigner1958distribution}. It states that if $X$ is a real symmetric matrix, then for any i.i.d.\footnote{The i.i.d. ensembles are also known as ``Wigner Class'' in random matrix theory literatures.} entry distribution, the asymptotic eigenvalue density of $\frac{1}{\sqrt N}X$ is given by a semicircle curve. With Wigner's semicircle law, it follows trivially that the asymptotic eigenvalue distribution of $A=\frac{1}{N}XX^\dagger$ is the MP distribution. However, this derivation of MP law from Wigner's semicircle law only applies to real symmetric $X$, \textit{i.e.} our case (5) in the list. Also, the original proof of Wigner's semicircle law~\cite{wigner1993characteristic,wigner1958distribution} did not use a diagrammatic method.

Diagrammatic proof of Wigner's semicircle law also exists in the literature~\cite{PhysRevE.49.2588,PhysRevE.53.1399,Zee:2003mt}. This diagrammatic method also applies to Hermitian $X$ (\textit{i.e.} our case (7)). When $X$ is real symmetric or Hermitian, one needs not to distinguish between $X$ and $X^\dagger$, hence only one kind of vertex is needed in the diagrams. Our work in this paper treats more general $X$ by using two types of vertices in the diagrams, and thus serves as a generalization of the diagrammatic method in~\cite{PhysRevE.49.2588,PhysRevE.53.1399,Zee:2003mt} to all the seven cases of $X$ listed.

Another paper particularly close to our work is~\cite{Feinberg:1996qq}, where a diagrammatic method was presented among some other methods to calculate the large $N$ asymptotic eigenvalue distribution of the matrix
\begin{equation}
H = \left( {\begin{array}{*{20}{c}}
0&{{X^\dag }}\\
X&0
\end{array}} \right) . \nonumber
\end{equation}
and of the matrix $A$ closely related to it, with $X$ being an arbitrary complex matrix (\textit{i.e.} our case (1)). However, an important difference is that the diagrammatic method presented in~\cite{Feinberg:1996qq} is limited to entry distribution being Gaussian. As to the proof of the universality for generic i.i.d. entry distributions, a large $N$ RG method was resorted. Remarkably, this RG picture can help better understanding the universality. But this approach is clearly not a \textit{direct diagrammatic method}. Actually, it was claimed in~\cite{Feinberg:1996qq} (the first paragraph of section 3) that it is rather difficult to develop a direct diagrammatic method for i.i.d. entry distributions beyond Gaussian. We believe the main difficulty is that a crucial prerequisite for a diagrammatic calculation is the notion of ``propagator'', which comes trivially under Gaussian ensemble, but not otherwise. In this paper, three subsections (Section~\ref{subsec:Group},~\ref{subsec:Key}, and~\ref{subsec:Associate}) are devoted to prove that the notion and use of ``propagator'' is legitimate because of a \textit{grouping by pairs} requirement (see Section~\ref{subsec:Associate} for details) under large $N$ limit. Therefore, we have overcome the difficulty claimed in~\cite{Feinberg:1996qq} and developed a direct diagrammatic method applicable to any i.i.d. entry distribution and any of the seven cases of $X$ listed.

The rest of this paper is organized as follows. We first give a precise statement of (a selected version of) the Marchenko-Pastur law in Section~\ref{sec:MP}. Then in Section~\ref{sec:Proof}, we present a detailed proof of MP law using Feynman diagrams. We generalize our proof to six types of restricted $X$ in Section~\ref{sec:Generalizing}. Section~\ref{sec:Conclusion} is our conclusion.

\section{Marchenko-Pastur Law \label{sec:MP}}

As mentioned in the introduction section, there are more than one versions of the MP law with the differences residing in how strong the premises are assumed. In this paper, we focus on its following version, which is sufficient for most large $N$ models in physics.

Let $X$ be a $M \times N$ random complex matrix, whose entries $X_{ij}$ are generated according to the following conditions:
\begin{align}
& \text{(1) independent, identical distribution (i.i.d.)} , \label{eq:con1} \\
& \text{(2) $\left\langle X_{ij} \right\rangle_X=0$, $\left\langle X_{ij}^2\right\rangle_X=0$, and $\left\langle \left| X_{ij} \right|^2\right\rangle_X=1$} , \label{eq:con2} \\
& \text{(3) $\left\langle {\left| {{X_{ij}}} \right|^{2 + \varepsilon}}\right\rangle_X < \infty$ for any $\varepsilon>0$} , \label{eq:con3}
\end{align}
where and throughout this paper, we use $\left\langle {\cal O} \right\rangle_X$ to denote the expectation value of a random variable ${\cal O}$ under the ensemble of $X$. Then construct an $M \times M$ hermitian matrix $A=\frac{1}{N}X{X^\dag}$, whose eigenvalues are denoted by $\lambda_k$, with $k=1,2,...,M$. Then the empirical distribution of these eigenvalues is defined as
\begin{equation}
{F_M}(x) \equiv \frac{1}{M}\sum\limits_{k = 1}^M {{I_{\{ {\lambda _k} \le x\} }}} ,
\end{equation}
where $I_{B}$ denotes the indicator of an event $B$:
\begin{equation}
{I_{\{ B\} }} = \left\{ \renewcommand\arraystretch{1.5} \begin{array}{ll}
 1 & \mbox{ } \mbox{ } \mbox{ if } B \mbox{ is true } \\
 0 & \mbox{ } \mbox{ } \mbox{ if } B \mbox{ is false } \\
 \end{array} \right.\
\end{equation}
Consider the limit $N \to \infty$. If the limit of the ratio $M/N$ is finite
\begin{equation}
b \equiv \mathop {\lim }\limits_{N \to \infty } M/N \in (0,\infty ) ,
\end{equation}
then $\left\langle F_M(x) \right\rangle_X \to F(x)$\footnote{The actual Marchenko-Pastur law states that $F_M(x)$ will converge to $F(x)$ in probability, namely that $\mathop{\lim}\limits_{N\to\infty} \text{Pr}(|F_M(x)-F(x)|>\epsilon)=0$ irrespective of $\epsilon>0$, where ``Pr'' stands for ``Probability''. This is a much stronger statement than $\left\langle F_M(x) \right\rangle_X \to F(x)$. But in this paper, we only prove the weaker version of the MP law.}, where $F(x)$ denotes the cumulative distribution function of the Marchenko-Pastur distribution whose density function is
\begin{eqnarray}
f(x) &=& \frac{1}{{2\pi }}\frac{{\sqrt {({x_2} - x)(x - {x_1})} }}{x}\frac{1}{b} \cdot {I_{\{ x \in ({x_1},{x_2})\} }} \nonumber \\
 && + (1 - \frac{1}{b})\delta (x) \cdot {I_{\{ b \in [1,\infty )\} }}\ , \label{eq:MPdensity}
\end{eqnarray}
with ${x_1} = {(1 - \sqrt b )^2}$ and ${x_2} = {(1 + \sqrt b )^2}$. In the special case of a square matrix $X$, namely $b=1$, this becomes
\begin{equation}
 f(x) = \frac{1}{2\pi} \sqrt{\frac{4}{x}-1} \cdot I_{x \in (0,4)} . \label{eq:MPdensityS}
\end{equation}

\section{Proof of Marchenko-Pastur Law with Feynman Diagrams \label{sec:Proof}}

\subsection{Stieltjes Transformation \label{subsec:Stieltjes}}

For a single matrix $X$ generated, the distribution density of eigenvalues is
\begin{equation}
{\rho _X}(E) = \frac{1}{M}\sum\limits_{k = 1}^M {\delta (E - {\lambda _k})} .
\end{equation}
Our goal is to compute its expectation
\begin{equation}
\rho (E) \equiv {\left\langle {{\rho _X}(E)} \right\rangle _X} = \int {dX \cdot {\rho _X}(E)} ,
\end{equation}
and prove that $\rho (E)$ approaches the MP density function (Eq.~\ref{eq:MPdensity}) as $N\to\infty$. Here we use $dX$ to denote the normalized measure of $X$:
\begin{equation}
 dX = \prod\limits_{ij} {g({X_{ij}})d{X_{ij}}}, \hspace{0.1cm} \int {dX} = 1 ,
\end{equation}
with $g(X_{ij})$ denoting the normalized distribution density of each $X_{ij}$.

Since $\rho(E)$ is not easy to compute directly, we make use of a method known in mathematics as ``Stieltjes transformation". That is, from the identity, where $x$ is a real variable
\begin{equation}
\delta (x) =  - \frac{1}{\pi }\mathop {\lim }\limits_{\varepsilon  \to {0^ + }} {\mathop{\rm Im}\nolimits} \frac{1}{{x + i\varepsilon }} ,
\end{equation}
we get
\begin{eqnarray}
\rho (E) &=& \int {\rho (x)\delta (E - x)dx} \nonumber \\
&=&  - \frac{1}{\pi }\mathop {\lim }\limits_{\varepsilon  \to {0^ + }} {\mathop{\rm Im}\nolimits} \int {\frac{{\rho (x)}}{{E + i\varepsilon  - x}}dx} . \label{eq:identity}
\end{eqnarray}
Thus for any distribution density function $\rho(x)$, we can define its Stieltjes transformation $G(z)$, a complex function as an integral over the support of $\rho(x)$
\begin{equation}
G(z) \equiv \int {\frac{{\rho (x)}}{{z - x}}dx} .
\end{equation}
Then according to Eq.~(\ref{eq:identity}), $\rho(x)$ can be obtained from the inverse formula
\begin{equation}
\rho (E) =  - \frac{1}{\pi }\mathop {\lim }\limits_{\varepsilon  \to {0^ + }} {\mathop{\rm Im}\nolimits} G(E + i\varepsilon ) . \label{eq:defrho}
\end{equation}

For our case, $G(z)$ can be computed as following
\begin{eqnarray}
 G(z) &=& \int {\frac{{\rho (x)}}{{z - x}}dx}  = \int {\frac{1}{{z - x}}{{\left\langle {{\rho _X}(x)} \right\rangle }_X}dx} \nonumber \\
 &=& {\left\langle {\frac{1}{M}\sum\limits_{k = 1}^M {\int {\frac{1}{{z - x}}\delta (x - {\lambda _k})dx} } } \right\rangle _X} \nonumber \\
 &=& {\left\langle {\frac{1}{M}\sum\limits_{k = 1}^M {\frac{1}{{z - {\lambda _k}}}} } \right\rangle _X} \nonumber \\
 &=& {\left\langle {\frac{1}{M}tr(\frac{1}{{z - A}})} \right\rangle _X} = \frac{1}{M}\frac{1}{z}tr\left[ {B(z)} \right] , \label{eq:defG}
\end{eqnarray}
where we have defined a matrix $B(z)$ as
\begin{eqnarray}
 B(z) &\equiv& {\left\langle {\frac{z}{{z - A}}} \right\rangle _X} \label{eq:defB} \\
 &=& {\left\langle {\sum\limits_{n = 0}^\infty  {{(\frac{A}{z})^n}} } \right\rangle _X} = {\left\langle {\sum\limits_{n = 0}^\infty  {{(\frac{1}{{zN}}X{X^\dag })^n}} } \right\rangle _X} . \label{eq:expansionB}
\end{eqnarray}
This expansion is a valid analytical form of $B(z)$ in the vicinity of $z=\infty$. We will compute $B(z)$ in this vicinity first, and then analytically continue it to the whole complex plane. Once $B(z)$ is obtained, $G(z)$ and $\rho(E)$ would follow immediately. The following several subsections are devoted to calculate this \textit{target function} $B(z)$ under the limit $N\to\infty$.

\subsection{Group into ``Boxes" \label{subsec:Group}}

Our target function is a sum of various terms, in which a typical $n$-term looks like
\begin{eqnarray}
 {B_{ij}}(z) &\supset& {(\frac{1}{{zN}})^n} \left\langle {\prod\limits_{p = 1}^n {{X_{{\alpha _p}{\beta _p}}}X_{{\beta _p}{\alpha _{p + 1}}}^\dag } }\right\rangle_X \nonumber \\
 &=& {(\frac{1}{{zN}})^n}{\left\langle {{X_{i{\beta _1}}}X_{{\beta _1}{\alpha _2}}^\dag  \cdots {X_{{\alpha _n}{\beta _n}}}X_{{\beta _n}j}^\dag } \right\rangle _X} , \label{eq:nterm}
\end{eqnarray}
with an identification $\alpha_1 \equiv i, \alpha_{n+1} \equiv j$ and a sum over all the dummy indices ${\alpha_2},{\alpha_3}, \cdots ,{\alpha_n}$ from $1$ to $M$, and ${\beta_1},{\beta_2}, \cdots ,{\beta_n}$ from $1$ to $N$.

As stated in the condition Eq.~(\ref{eq:con1}), different elements $X_{ij}$ are independent. Therefore any such $n$-term expectation can be factorized
\begin{eqnarray}
 && {\left\langle {{X_{i{\beta _1}}}X_{{\beta _1}{\alpha _2}}^\dag  \cdots {X_{{\alpha _n}{\beta _n}}}X_{{\beta _n}j}^\dag } \right\rangle _X} \nonumber \\
 &=& {\left\langle {f({X_{{k_1}{l_1}}})} \right\rangle _X}{\left\langle {f({X_{{k_2}{l_2}}})} \right\rangle _X} \cdots ,
\end{eqnarray}
where each individual expectation ${\left\langle {f(X_{kl})} \right\rangle _X}$ contains only $X_{kl}$ and its complex conjugate $X_{kl}^*$
\begin{equation}
{\left\langle {f({X_{kl}})} \right\rangle _X} = {\left\langle {{{({X_{kl}})}^{{m_1}}}{{(X_{kl}^*)}^{{m_2}}}} \right\rangle _X} .
\end{equation}
Namely that the independence among elements allows us to group same $X_{kl}$ (and the complex conjugate) together into one factor. For future convenience, let us call each such factor a ``box".

Now in evaluating the $n$-term (Eq.~\ref{eq:nterm}), the sum over the dummy indices $\alpha$'s and $\beta$'s can be decomposed into two steps. (1) There are many ways to group the $2n$ elements into boxes. We need to sum over all possible grouping configurations. (2) Under each grouping configuration, all $\alpha$'s within the same box are forced equal, so are all $\beta$'s, thus some dummy indices are tied to others and hence no longer free to sum. But generically there are still some \textit{free dummy indices} remaining, which needs to be summed. In summary, this decomposition of sum can be expressed as
\begin{equation}
\sum\limits_{\alpha,\beta}=\sum\limits_{\text{grouping configurations}} \hspace{0.2cm} \sum\limits_{\text{free dummy indices}} . \label{eq:sumdecompose}
\end{equation}

\subsection{Key Statement in Power Counting of $N$ \label{subsec:Key}}

To evaluate an $n$-term (Eq.~\ref{eq:nterm}) under $N\to\infty$, an efficient way to count the power of $N$ is definitely crucial. The suppression factor $\frac{1}{N^n}$ in front of Eq.~(\ref{eq:nterm}) contributes a factor $N^{-n}$. On the other hand, with the sum decomposition Eq.~\ref{eq:sumdecompose}, under each grouping configuration, summing over each free dummy index will give us one power of $N$: $\sum\limits_{\alpha=1}^{M} \delta_{\alpha\alpha}=M=bN, \sum\limits_{\beta=1}^{N} \delta_{\beta\beta}=N$. From this competition, we end up with a factor $N^{-(n-n_f)}$, where $n_f$ denotes the total number of free dummy indices under a given grouping configuration. There are $2n-1$ dummy indices in total: $\alpha_2,\cdots,\alpha_n$, $\beta_1,\cdots,\beta_n$, but not all of them are free, because within each box, the different $\alpha$'s and $\beta$'s are forced into same value respectively. Apparently, $n_f$ largely depends on the grouping configuration.

To study how $n_f$ depends on the grouping configuration, we resort to some graphical analysis. First, let us draw a map to represent each grouping configuration, where each grouping box is drawn as an isolated island. We notice that in the sequence of Eq.~(\ref{eq:nterm})
\begin{equation}
{X_{i{\beta _1}}}X_{{\beta _1}{\alpha _2}}^\dag  \cdots {X_{{\alpha _n}{\beta _n}}}X_{{\beta _n}j}^\dag , \label{eq:sequence}
\end{equation}
every dummy index appears twice, i.e. in pair. Each such dummy index pair could be grouped into the same box, or two different boxes. If any two boxes share a dummy index pair, let us connect those two islands by a ``bridge" on the map. So every grouping configuration is described by a map of boxes (or islands) and a number of bridges connecting them.

Suppose that there are $m$ boxes fully connected by bridges. Within each box, after all $\alpha$'s and $\beta$'s are identified respectively, we are left with at most $2$ free dummy indices. Then collecting all the $m$ boxes, we get at most $2m$ free indices. But to connect all these $m$ boxes, we need at lease $m-1$ bridges. If we remove all the redundant bridges and thus chop the map into a tree map, then each of the remaining $m-1$ bridges would effectively reduce one free index out of the $2m$. Thus we have arrived at our \textit{key statement}: \\

\textbf{If $m$ boxes are fully connected by bridges, then we can get at most $m+1$ free dummy indices out of them.}\\

Clearly, the whole sequence Eq.~(\ref{eq:sequence}) is fully connected by dummy-index bridges, so we can apply this key statement to it. Suppose there are $n_b$ boxes in total for a grouping configuration, then we get $n_f \le n_b+1$. However, this upper bound is obtained by counting $i$ and $j$ also as dummy indices. But they are not. So we need to further subtract $1$, if $i$ and $j$ are already identified by the grouping; or subtract $2$ if they are not. To sum up, we get
\begin{equation}
{n_f} \le \left\{ \renewcommand\arraystretch{1.5} \begin{array}{ll}
 n_b & \mbox{ $i$, $j$ identified by the grouping} \\
 n_b-1 & \mbox{ $i$, $j$ not identified by the grouping} \\
 \end{array} \right.\ \label{eq:nfresult}
\end{equation}

\subsection{Association with Feynman Diagrams \label{subsec:Associate}}

With the result Eq.~(\ref{eq:nfresult}), we are ready to study what kinds of grouping configurations can give nonzero contribution. Due to the condition Eq.~(\ref{eq:con2}), each box needs to contain at least two elements in order not to vanish. But there are in total only $2n$ elements in an $n$-term. So if any box has more than two elements, then the total number of boxes $n_b$ must be less than $n$. Consequently, $n_f \le n_b < n$ and the grouping configuration is suppressed by a factor $N^{-(n-n_f)}$. Due to condition Eq.~(\ref{eq:con3}), the coefficient multiplying this factor must be finite. Thus the contribution from this kind of grouping configuration vanishes under $N\to\infty$. To sum up, only grouping the elements by pairs can give nonzero contributions. We can call each such grouped pair a ``contraction''. We emphasize that it is this \textit{grouping by pairs} requirement that justifies the notion and the use of ``contraction'', which in turn makes a diagrammatic approach possible.

Here is also a bonus result: Even under pair grouping configurations where $n_b=n$, $i$ and $j$ must be identified by the grouping in order to get large enough $n_f$ (see Eq.~\ref{eq:nfresult}). So $i$ and $j$ must be equal, which means that the matrix $B(z)$ must be diagonal under $N\to\infty$.

\begin{figure}[t]
 \centering
 \includegraphics[height=1.8cm]{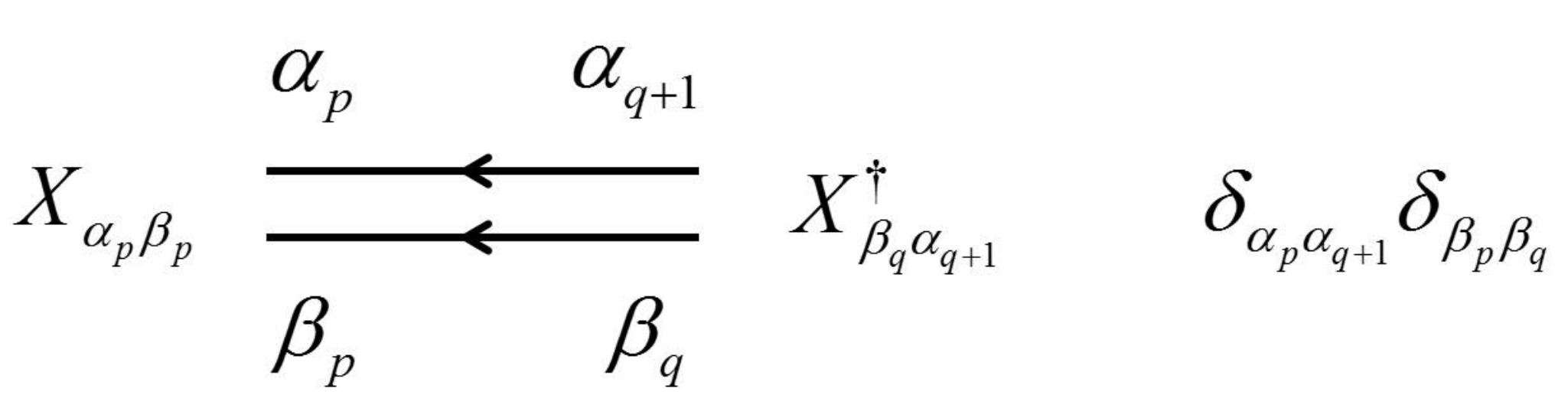}
 \caption{Feynman rule propagator}\label{fig:propagator}
\end{figure}

According to the condition Eq.~(\ref{eq:con2}), only contracting $X$ and $X^\dag$ can be nonzero
\begin{equation}
{\left\langle {{X_{ij}}X_{kl}^\dag } \right\rangle _X} = {\left\langle {{X_{ij}}X_{lk}^*} \right\rangle _X} = {\delta _{il}}{\delta _{jk}} . \label{eq:propagator}
\end{equation}
Let us call this contraction ``propagator", which corresponds to the Feynman rule shown in Fig.~\ref{fig:propagator}.

\begin{figure}[t]
 \subfigure{
 \centering
 \includegraphics[height=1.8cm]{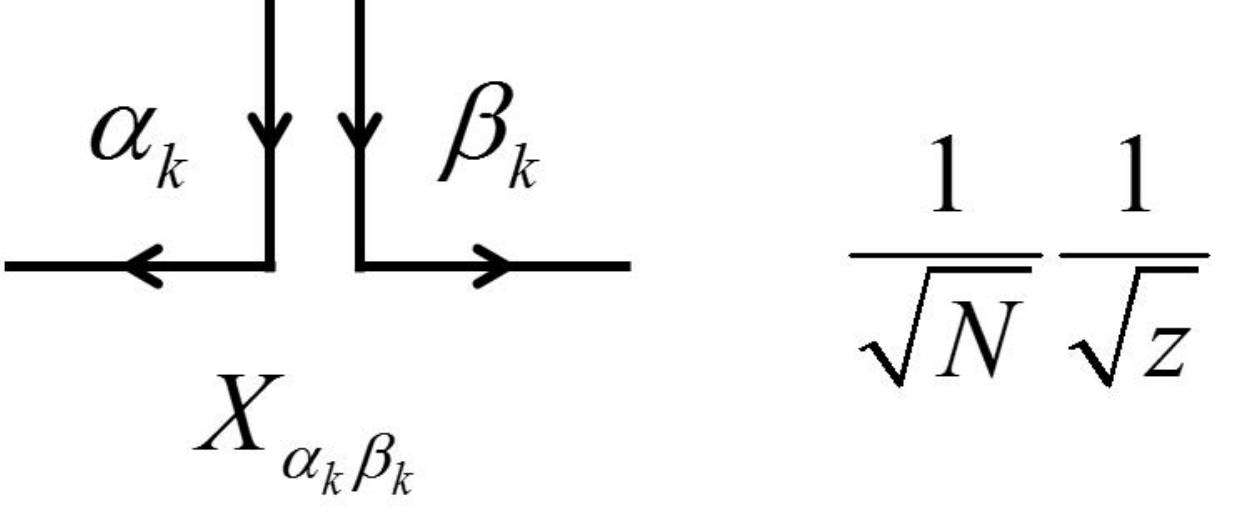}
 }
 \subfigure{
 \centering
 \includegraphics[height=1.8cm]{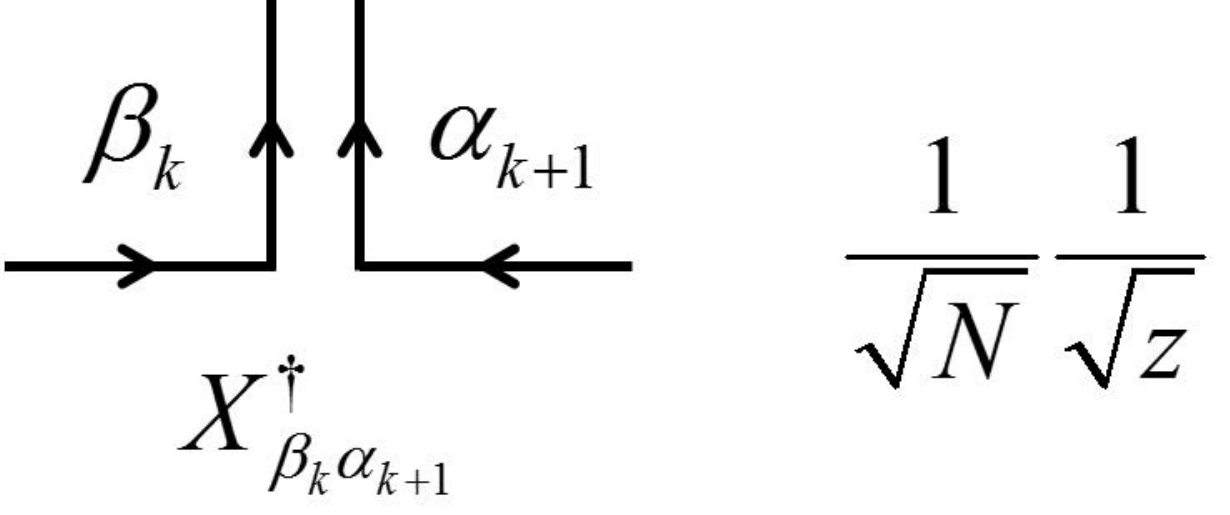}
 }
 \caption{Feynman rule vertices}\label{fig:vertices}
\end{figure}

\begin{figure*}[t]
 \centering
 \includegraphics[height=2cm]{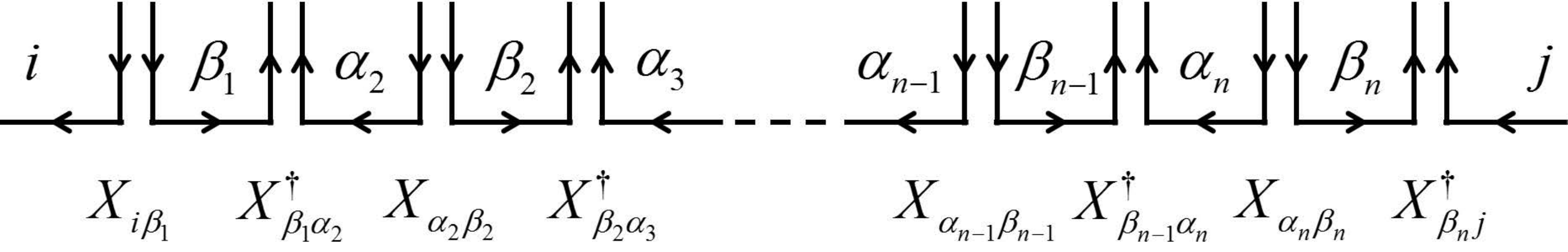}
 \caption{Feynman diagrams of $n$-term (before specifying contraction structure).}\label{fig:nterm}
\end{figure*}

\begin{figure*}[t]
 \subfigure[$\hspace{0.2cm} n=0$ term]{
 \begin{minipage}[b]{0.3\textwidth}
 \centering
 \includegraphics[height=1cm]{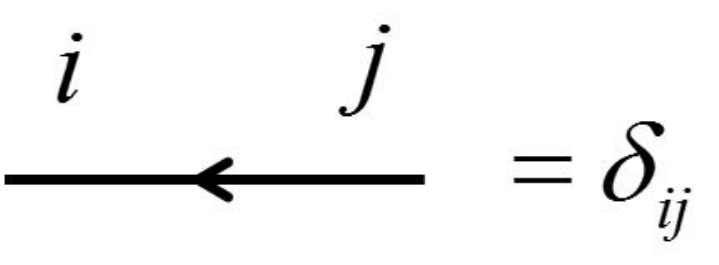}\label{fig:example0}
 \vspace{0.5cm}
 \end{minipage}}
 \subfigure[$\hspace{0.2cm} n=1$ term]{
 \centering
 \includegraphics[height=2.2cm]{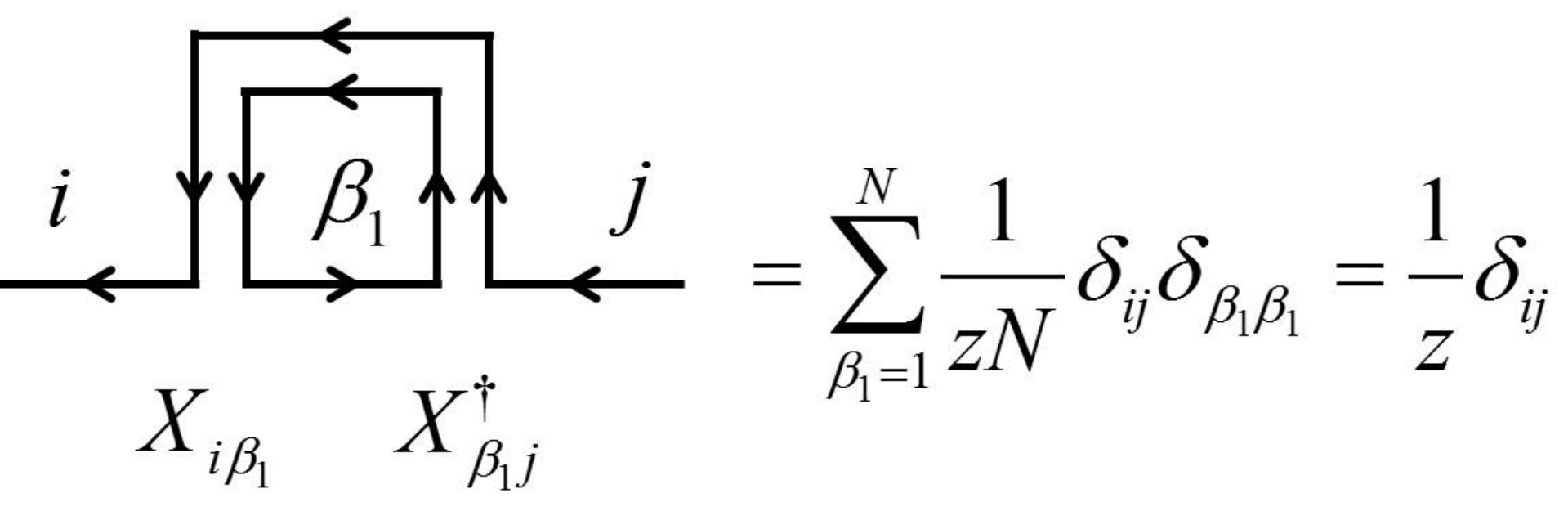}\label{fig:example1}
 }\vspace{0.5cm}
 \subfigure[$\hspace{0.2cm} n=2$ case 1]{
 \centering
 \includegraphics[height=2.3cm]{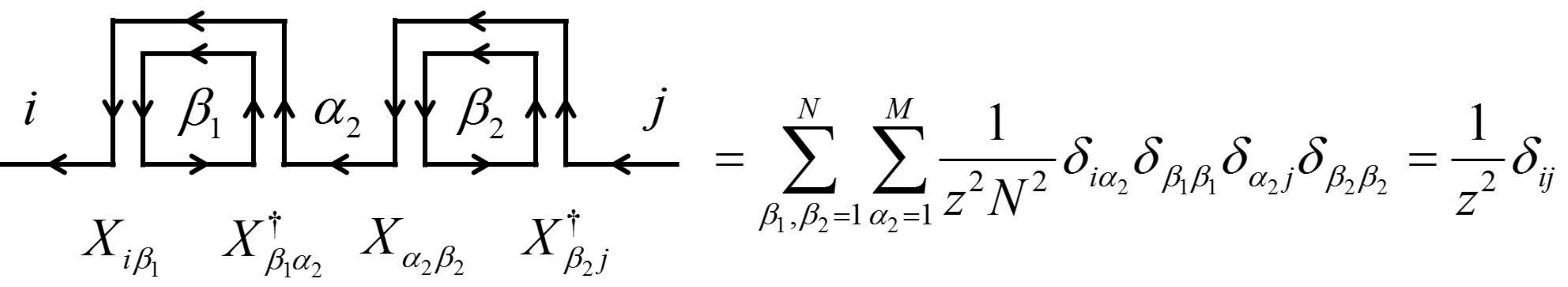}\label{fig:example21}
 }\vspace{0.5cm}
 \subfigure[$\hspace{0.2cm} n=2$ case 2]{
 \centering
 \includegraphics[height=2.8cm]{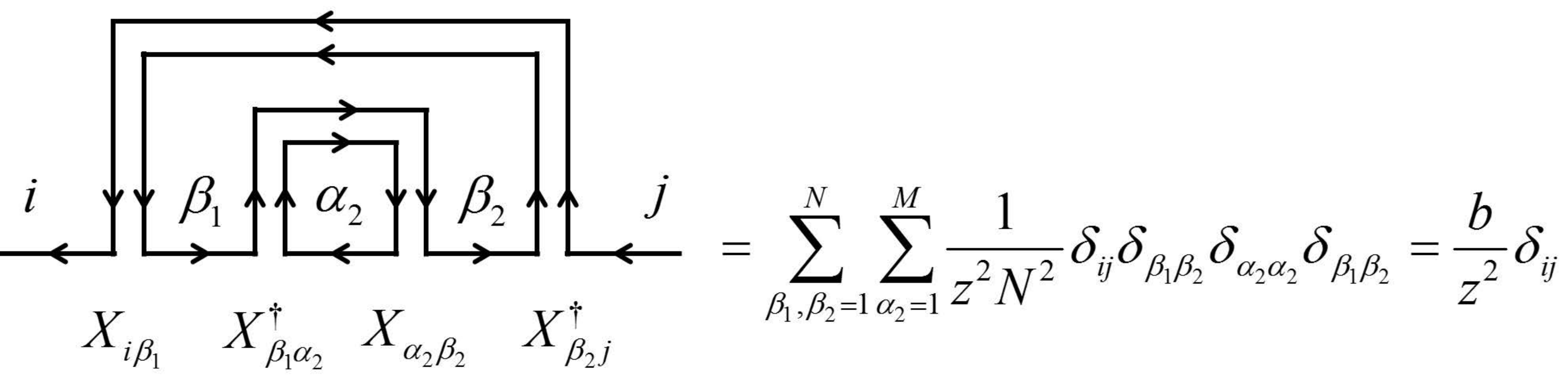}\label{fig:example22}
 }
 \caption{Feynman diagram examples.}\label{fig:examples}
\end{figure*}

Our goal is to evaluate the target function $B(z)$ (Eq.~\ref{eq:expansionB}). Since each matrix element has two indices, and each pair of $XX^\dagger$ always comes with a factor $\frac{1}{zN}$, we are naturally led to the Feynman rules of vertices shown in Fig.~\ref{fig:vertices}. The arrow flow is used to distinguish $X$ from $X^\dagger$. This is necessary because our $X$ is generically not hermitian.
Then a typical $n$-term (Eq.~\ref{eq:nterm}) can be calculated by summing over Feynman diagrams corresponding to all the possible contraction structures of Fig.~\ref{fig:nterm}.
To give a few examples, we enumerate all nonzero diagrams contributing to $n=0$, $n=1$ and $n=2$ terms in Fig.~\ref{fig:examples}.

\subsection{Simplification: Planar Diagrams only for $N\to\infty$ \label{subsec:Simplification}}

Now we have developed a diagrammatic way of evaluating $B_{ij}(z)$ as described by Fig.~\ref{fig:nterm}, which is well organized and quite routine. But the actual calculation is still rather complicated, because there are so many ways of contracting the vertices. Large $N$ limit, however, brings us another great simplification: Any diagram with crossed contractions will vanish under $N\to\infty$. This means that we only need to consider the type of contraction shown in the first line of the following, but not that kind shown in the second line.
\begin{align}
\contraction[2ex]{\cdots}{X}{_{\alpha_2 \beta_2} X_{\beta_2 \alpha_3}^\dag X_{\alpha_3 \beta_3}}{X}
\contraction{\cdots X_{\alpha_2 \beta_2}}{X}{_{\beta_2 \alpha_3}^\dag}{X}
\cdots X_{\alpha_2 \beta_2} X_{\beta_2 \alpha_3}^\dag X_{\alpha_3 \beta_3} X_{\beta_3 \alpha_4}^\dag X_{\alpha_4 \beta_4} X_{\beta_4 \alpha_5}^\dag \cdots \nonumber
\end{align}
\begin{align}
\contraction[2ex]{\cdots}{X}{_{\alpha_2 \beta_2} X_{\beta_2 \alpha_3}^\dag X_{\alpha_3 \beta_3}}{X}
\contraction{\cdots X_{\alpha_2 \beta_2}}{X}{_{\beta_2 \alpha_3}^\dag X_{\alpha_3 \beta_3} X_{\beta_3 \alpha_4}^\dag}{X}
\cdots X_{\alpha_2 \beta_2} X_{\beta_2 \alpha_3}^\dag X_{\alpha_3 \beta_3} X_{\beta_3 \alpha_4}^\dag X_{\alpha_4 \beta_4} X_{\beta_4 \alpha_5}^\dag \cdots \nonumber
\end{align}
Once crossed contractions are forbidden, all the propagators can only form two types of structures: ``side by side" as in the example of Fig.~\ref{fig:example21} or ``nesting" as in Fig.~\ref{fig:example22}. A combination of these two types gives us a general ``planar" diagram. Only planar diagrams have nonzero contributions under $N\to\infty$.

This requirement also follows from our key statement. Assume that we have a contraction jumping $k$ couples of elements:
\begin{align}
\contraction{\cdots X_{\beta_{p-1}\alpha_p}^\dag}{X}{_{\alpha_p \beta_p} X_{\beta_p \alpha_{p+1}}^\dag \cdots X_{\alpha_q \beta_q}}{X}
\cdots X_{\beta_{p-1}\alpha_p}^\dag X_{\alpha_p \beta_p} X_{\beta_p \alpha_{p+1}}^\dag \cdots X_{\alpha_q \beta_q} X_{\beta_q \alpha_{q+1}}^\dag X_{\alpha_{q+1}\beta_{q+1}} \cdots \nonumber
\end{align}
with $q=p+k$. This contraction identifies $\alpha_{q+1}$ with $\alpha_p$ and $\beta_q$ with $\beta_p$. After summing over these non-free dummy indices $\alpha_{q+1}$ and $\beta_q$, we get the result proportional to (with a finite coefficient)
\begin{align}
\cdots X_{{\beta _{p - 1}}{\alpha _p}}^\dag  \cdot X_{{\beta _p}{\alpha _{p + 1}}}^\dag  \cdots {X_{{\alpha _q}{\beta _p}}} \cdot {X_{{\alpha _p}{\beta _{q + 1}}}} \cdots \label{eq:contractedsequence}
\end{align}
Now we are only left with $n-1$ couples of $X$ and $X^\dag$. With the number of boxes $n_b=n-1$ then, it seems impossible to make $n_f=n$, according to our previous result Eq.~(\ref{eq:nfresult}). However, by a careful look at the new sequence Eq.~(\ref{eq:contractedsequence}), we realize that it is no longer guaranteed to be fully connected by bridges. Instead, it consists of two parts: inside the contraction and outside the contraction, each part fully connected. So as long as we do not group any element inside with any element outside into one box (i.e. no crossed contraction!), we can only apply our key statement to each part separately. In this case, we are just lucky enough to save it: we can get up to $k+1$ free dummy indices from the inside part and $n-1-k$ from the outside part. Together, we can still make $n_f=n$. On the other hand, if we do make a contraction crossed with the first one, then the divided two parts are reconnected through this contraction box and Eq.~(\ref{eq:nfresult}) can be applied to the whole sequence Eq.~(\ref{eq:contractedsequence}): $n_f \le n_b=n-1<n$. Therefore diagram with crossed contractions will vanish under $N\to\infty$.

\subsection{Diagrammatic Calculation for $N\to\infty$ \label{subsec:Diagrammatic}}

\begin{figure}[t]
 \centering
 \includegraphics[height=4.5cm]{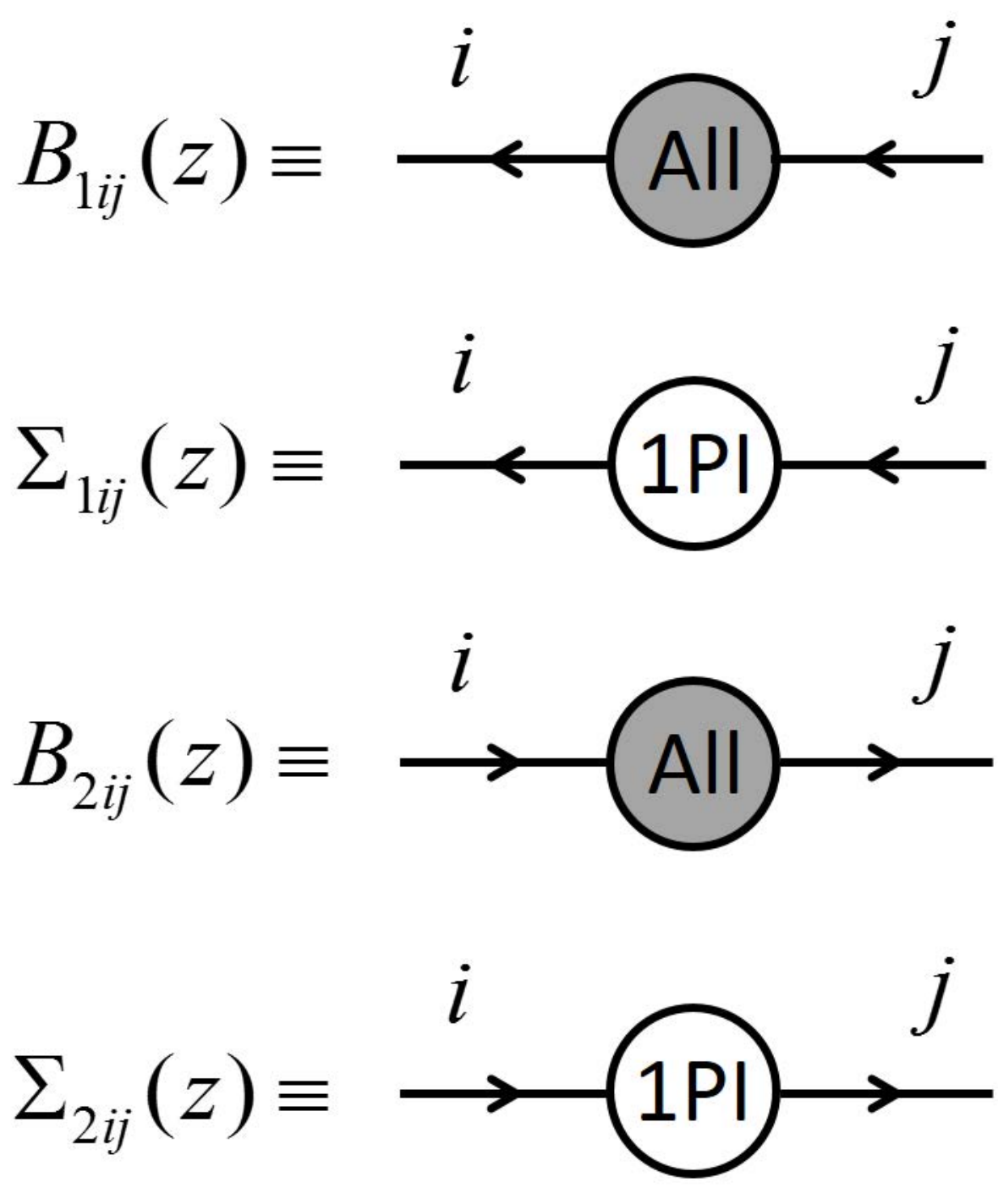}
 \caption{Auxiliary functions defined in terms of planar Feynman diagrams.}\label{fig:diagramsdef}
\end{figure}

Now we are finally ready to calculate our target function $B_{ij}(z)$ by summing over all the planar Feynman diagrams formed from Fig.~\ref{fig:nterm}. For convenience, let us define four functions as shown in Fig.~\ref{fig:diagramsdef}, two 1PI (1 Particle Irreducible) functions $\Sigma_{1ij}(z),\hspace{0.1cm} \Sigma_{2ij}(z)$, and two two-point functions $B_{1ij}(z),\hspace{0.1cm} B_{2ij}(z)$. Here $B_{1ij}$ is nothing but our target function $B_{ij}(z)=B_{1ij}(z)$.

\begin{figure}[t]
 \subfigure[\hspace{0.2cm} $\Sigma_{1ij}(z)$]{
 \centering
 \includegraphics[height=2.5cm]{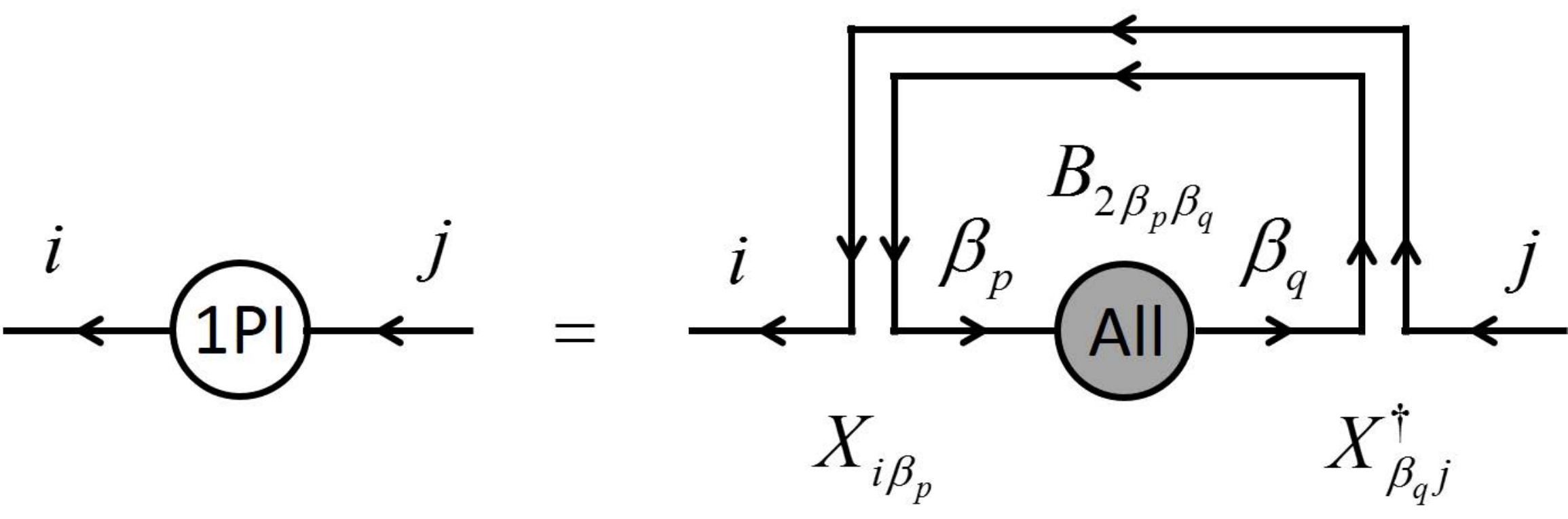}\label{fig:sigma1}
 }\vspace{0.5cm}
 \subfigure[\hspace{0.2cm} $\Sigma_{2ij}(z)$]{
 \centering
 \includegraphics[height=2.5cm]{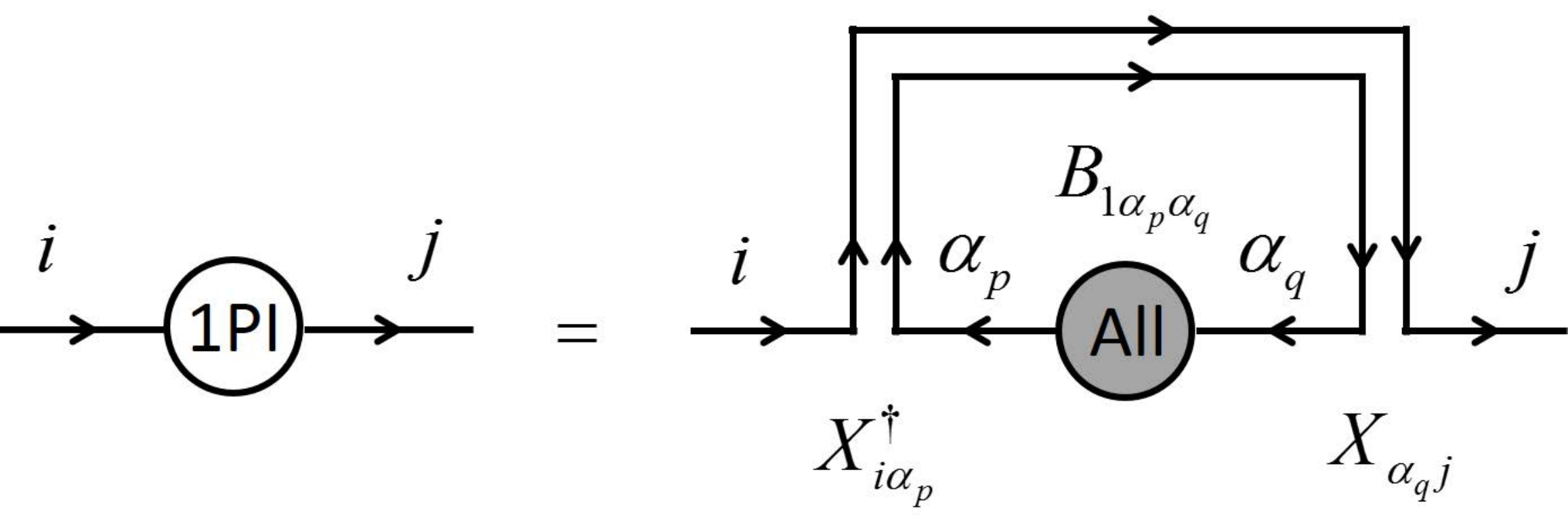}\label{fig:sigma2}
 }
 \caption{1PI diagrams (a) $\Sigma_{1ij}(z)$, and (b) $\Sigma_{2ij}(z)$, in terms of two point full diagrams}
\end{figure}

First, let us study the 1PI functions. $\Sigma_{1ij}(z)$ sums over all the 1PI planar diagrams with external single arrows pointing to the left. Each 1PI planar diagram must have a double-line contraction coating it at the most outside, with nested inside anything. Clearly, the sum of the nested part gives nothing but $B_2(z)$. So we get a relation as shown in Fig.~\ref{fig:sigma1}:
\begin{eqnarray}
 {\Sigma _{1ij}} &=& \sum\limits_{{\beta _p},{\beta _q} = 1}^N {\frac{1}{{zN}}{\delta _{ij}}{\delta _{{\beta _p}{\beta _q}}}{B_{2{\beta _p}{\beta _q}}}} \nonumber \\
 &=& \left( {\frac{1}{{zN}}\sum\limits_{{\beta _p} = 1}^N {{B_{2{\beta _p}{\beta _p}}}} } \right){\delta _{ij}} \equiv {\Sigma _1}{\delta _{ij}} . \label{eq:sigma1}
\end{eqnarray}
We see that $\Sigma_{1ij}(z)$ is proportional to the identity matrix. There is a similar relation for $\Sigma_{2ij}(z)$ (as shown in Fig.~\ref{fig:sigma2}), which is also proportional to identity matrix:
\begin{eqnarray}
 {\Sigma _{2ij}} &=& \sum\limits_{{\alpha _p},{\alpha _q} = 1}^M {\frac{1}{{zN}}{\delta _{ij}}{\delta _{{\alpha _p}{\alpha _q}}}{B_{1{\alpha _p}{\alpha _q}}}} \nonumber \\
 &=& \left( {\frac{1}{{zN}}\sum\limits_{{\alpha _p} = 1}^M {{B_{1{\alpha _p}{\alpha _p}}}} } \right){\delta _{ij}} \equiv {\Sigma _2}{\delta _{ij}} . \label{eq:sigma2}
\end{eqnarray}

Now let us turn to the two point functions $B_{1ij}(z)$ and $B_{2ij}(z)$. Same as in computing a two-point correlation function in QFT, all the diagrams contributing to $B_{1ij}(z)$ ($B_{2ij}(z)$) can be organized into a geometric series of the 1PI functions $\Sigma_{1ij}(z)$ ($\Sigma_{2ij}(z)$). Since both $\Sigma_{1ij}(z)$ and $\Sigma_{2ij}(z)$ are proportional to identity matrix, $B_{1ij}(z)$ and $B_{2ij}(z)$ are also proportional to identity matrix:
\begin{eqnarray}
{B_{1ij}} &=& {\delta _{ij}} + {\Sigma _{1ij}} + {\Sigma _{1i\alpha }}{\Sigma _{1\alpha j}} +  \ldots \nonumber \\
 &=& (1 + {\Sigma _1} + \Sigma _1^2 +  \ldots ){\delta _{ij}} = \frac{1}{{1 - {\Sigma _1}}}{\delta _{ij}} \nonumber \\
 &\equiv& {B_1}{\delta _{ij}} , \\
{B_{2ij}} &=& {\delta _{ij}} + {\Sigma _{2ij}} + {\Sigma _{2i\beta }}{\Sigma _{2\beta j}} +  \ldots \nonumber \\
 &=& (1 + {\Sigma _2} + \Sigma _2^2 +  \ldots ){\delta _{ij}} = \frac{1}{{1 - {\Sigma _2}}}{\delta _{ij}} \nonumber \\
 &\equiv& {B_2}{\delta _{ij}} .
\end{eqnarray}
This confirms our bonus result in subsection~\ref{subsec:Associate} that $B_{1ij}$ and $B_{2ij}$ have to be diagonal. Going back to Eq.~(\ref{eq:sigma1}) and Eq.~(\ref{eq:sigma2}), we get
\begin{eqnarray}
 {\Sigma _1} &=& \frac{1}{{zN}}\sum\limits_{{\beta _p} = 1}^N {{B_{2{\beta _p}{\beta _p}}}}  = \frac{{{B_2}}}{{zN}}\sum\limits_{{\beta _p} = 1}^N {{\delta _{{\beta _p}{\beta _p}}}}  = \frac{1}{z}{B_2} , \\
 {\Sigma _2} &=& \frac{1}{{zN}}\sum\limits_{{\alpha _p} = 1}^M {{B_{1{\alpha _p}{\alpha _p}}}}  = \frac{{{B_1}}}{{zN}}\sum\limits_{{\alpha _p} = 1}^M {{\delta _{{\alpha _p}{\alpha _p}}}}  = \frac{b}{z}{B_1} .
\end{eqnarray}

Combining the work above, we get the following equation set
\begin{equation}
\left\{ \renewcommand\arraystretch{2.2} \begin{array}{l}
{B_1} = \dfrac{1}{{1 - {\Sigma _1}}}\\
{B_2} = \dfrac{1}{{1 - {\Sigma _2}}}\\
{\Sigma _1} = \dfrac{1}{z}{B_2}\\
{\Sigma _2} = \dfrac{b}{z}{B_1}
\end{array} \right.
\end{equation}
Because we are eventually interested in $B_{ij}(z)=B_{1ij}(z)=B_1(z)\delta_{ij}$, we eliminate the other three variables and get the equation of $B_1(z)$:
\begin{equation}
b B_1^2 - \left[ {z - (1 - b)} \right]{B_1} + z = 0 . \label{eq:Bequation}
\end{equation}
Solving this and plugging it into Eq.~(\ref{eq:defG}) and Eq.~(\ref{eq:defrho}), we get the result of $\rho(E)$
\begin{eqnarray}
 \rho (E) &=&  - \frac{1}{\pi }\mathop {\lim }\limits_{\varepsilon  \to {0^ + }} {\mathop{\rm Im}\nolimits} G(E + i\varepsilon ) \nonumber \\
 &=& \frac{1}{{2\pi }}\frac{1}{b}\frac{{\sqrt {({x_2} - E)(E - {x_1})} }}{E}{I_{\{ E \in ({x_1},{x_2})\} }} \nonumber \\
 && + (1 - \frac{1}{b})\delta (E){I_{\{ b \ge 1\} }} , \label{eq:rhoresult}
\end{eqnarray}
which is exactly what we want to prove (Eq.~\ref{eq:MPdensity}). Note that there are two solutions for Eq.~(\ref{eq:Bequation}), and one needs be cautious while choosing the root and taking the limit. It is a straight forward but slightly tedious procedure. To keep this paper self contained, we include this procedure as the appendix.

\section{Generalizing to Other Cases of $X$}\label{sec:Generalizing}

As mentioned in the introduction section, our direct diagrammatic approach can be readily generalized to six types of restricted $X$. This makes a total list of seven cases of $X$ which we reproduce here for convenience:
\begin{eqnarray}
&&\text{(1) Complex arbitrary} \nonumber \\
&&\text{(2) Complex symmetric} \nonumber \\
&&\text{(3) Complex antisymmetric} \nonumber \\
&&\text{(4) Real arbitrary} \nonumber \\
&&\text{(5) Real symmetric} \nonumber \\
&&\text{(6) Real antisymmetric} \nonumber \\
&&\text{(7) Hermitian} \nonumber
\end{eqnarray}
Many of these cases are interesting in physics models. For example, in the case of large $N$ analysis of neutrino anarchy, the Majorana mass matrix is complex symmetric (case (2) above).

It is understood that for the cases (2)-(7) above, the conditions (Eq.~\ref{eq:con1}-Eq.~\ref{eq:con3}) on the random entries of $X$ should be modified accordingly. First, in the condition Eq.~\eqref{eq:con1}, ``independent'' should be understood as only among the \textit{free entries} of $X$. For cases (1) and (4), $X$ has $M\times N$ free entries. Other cases require that $M=N$ ($b=1$). For cases (2), (5), and (7), $X$ has $N(N+1)/2$ free entries.\footnote{For case (2), all of these free entries are complex valued. For case (5), all of these free entries are real valued. For case (7), $N(N-1)/2$ of these free entries are complex valued, and $N$ of these free entries are real valued.} For cases (3) and (6), $X$ has $N(N-1)/2$ free entries. Second, the condition Eq.~\eqref{eq:con2} is specifically for complex valued entries of $X$. In certain cases above, all or part of the entries of $X$ are required to be real valued. For real valued $X_{ij}$, the condition Eq.~\eqref{eq:con2} should be replaced by the following
\begin{equation}
\left\langle X_{ij} \right\rangle_X=0,\hspace{2mm} \left\langle X_{ij}^2\right\rangle_X=1 . \nonumber
\end{equation}

To see that our whole analysis through Section~\ref{sec:Proof} still works for the other six cases of $X$, we need a few observations. First, it is clear that the \textit{grouping by pairs} requirement (see the first paragraph of Section~\ref{subsec:Associate}) holds for all the seven cases above. As we emphasized before, this requirement justifies the notion and the use of ``contraction'' and makes a diagrammatic approach possible. Second, one may worry that for cases (4)-(7), $\left\langle X_{ij} X_{kl}\right\rangle_X$ can be nonzero, namely that $X$ can contract with not only $X^\dagger$ but also $X$. This will not be a problem, because when an $X$ contracts with another $X$ (or an $X^\dagger$ contracts with another $X^\dagger$) in Eq.~\eqref{eq:sequence}, an odd number of entries of $X$ are left inside this contraction, which makes a crossed contraction inevitable. And from Section~\ref{subsec:Simplification}, we know that terms with crossed contractions vanish under $N\to\infty$. So we still only need to consider the type of contraction $\left\langle X_{ij} X_{kl}^\dagger\right\rangle_X$, namely the type of propagator shown in Fig.~\ref{fig:propagator}, even in cases (4)-(7). However, for cases (2)-(7), the value of the propagator could be different from Eq.~\eqref{eq:propagator} or that shown in Fig.~\ref{fig:propagator}. Therefore, a final observation needed is how Eq.~\eqref{eq:propagator} is changed and how that affects the calculations through Section~\ref{sec:Proof}. It is easy to see that for the seven cases of $X$ above, Eq.~\eqref{eq:propagator} should be modified into three values
\begin{eqnarray}
{\left\langle {{X_{ij}}X_{kl}^\dag } \right\rangle _{X,\text{ arbitrary}}} &=& \delta_{il} \delta_{jk} , \label{eq:prop} \\
{\left\langle {{X_{ij}}X_{kl}^\dag } \right\rangle _{X,\text{ symmetric}}} &=& \delta_{il} \delta_{jk} + \delta_{ik} \delta_{jl}(1-\delta_{ij}) , \label{eq:propsym} \\
{\left\langle {{X_{ij}}X_{kl}^\dag } \right\rangle _{X,\text{ antisymmetric}}} &=& \delta_{il} \delta_{jk} - \delta_{ik} \delta_{jl} . \label{eq:propasym}
\end{eqnarray}
In the above, Eq.~\eqref{eq:prop} applies to cases (1), (4), and (7); Eq.~\eqref{eq:propsym} applies to cases (2) and (5); and Eq.~\eqref{eq:propasym} applies to cases (3) and (6). Clearly, the modifications of Eq.~\eqref{eq:propagator} are additional terms proportional to $\delta_{ik}\delta_{jl}$. This modification could affect our calculations in Section~\ref{sec:Proof} only through Eq.~\eqref{eq:sigma1}) and \eqref{eq:sigma2}. However, we can easily see that the additional terms in these two equations vanish under $N\to\infty$
\begin{eqnarray}
 {\Sigma _{1ij\text{, additional}}}(z) &\propto& \sum\limits_{{\beta _p},{\beta _q} = 1}^N {\frac{1}{{zN}}{\delta _{i{\beta _q}}}{\delta _{{\beta _p}j}}{B_{2{\beta _p}{\beta _q}}}} \nonumber \\
 &=& \frac{1}{{zN}}{B_{2ji}} \to 0 , \nonumber \\
 {\Sigma _{2ij\text{, additional}}}(z) &\propto& \sum\limits_{{\alpha _p},{\alpha _q} = 1}^M {\frac{1}{{zN}}{\delta _{i{\alpha _q}}}{\delta _{{\alpha _p}j}}{B_{1{\alpha _p}{\alpha _q}}}} \nonumber \\
 &=& \frac{1}{{zN}}{B_{1ji}} \to 0 . \nonumber
\end{eqnarray}
Therefore, our diagrammatic proof presented in Section~\ref{sec:Proof} holds for all the seven cases of $X$.

\section{Conclusions \label{sec:Conclusion}}

Method with large $N$ random matrices is greatly used in various of theoretical models. Marchenko-Pastur law is a useful theorem for eigenvalue distribution of large $N$ random matrices. We present a direct diagrammatic approach of calculating the Marchenko-Pastur distribution. We also show that our direct diagrammatic approach can be readily generalized to six types of restricted random matrices.

\begin{acknowledgments}
We thank Joshua Feinberg and Anthony Zee for bringing our attentions to useful literatures on diagrammatic methods in random matrix theory. This work was supported in part by the U.S. DOE under Contract DE-AC03-76SF00098, and in part by the NSF under grant PHY-1002399. The work by H.M. was also supported in part by the JSPS grant (C) 23540289, in part by the FIRST program Subaru Measurements of Images and Redshifts (SuMIRe), CSTP, Japan, and by WPI, MEXT, Japan.
\end{acknowledgments}

\appendix*
\section{Root Selection}

Let us start with Eq.~(\ref{eq:Bequation}):
\begin{equation}
b B_1^2 - \left[ {z - (1 - b)} \right]{B_1} + z = 0 . \label{eq:BequationA}
\end{equation}
This equation gives us two analytical solutions, which for the moment, we formally write as
\begin{equation}
{B_1}(z) = \frac{{z - (1 - b) - r(z)}}{{2b}} , \label{eq:Bsolution}
\end{equation}
where we have used $r(z)$ to denote the square root
\begin{equation}
r(z)\equiv{\left\{ {{{\left[ {z - (1 - b)} \right]}^2} - 4bz} \right\}^{\frac{1}{2}}} ,
\end{equation}
and put in by hand a minus sign in front of it, just for future convenience. The selection of root is still undone until we specify the branch of this multi-value function $r(z)$. The expression of $G(z)$ follows
\begin{eqnarray}
 G(z) &=& \frac{1}{M}\frac{1}{z}tr\left[ {B(z)} \right] = \frac{1}{z}{B_1}(z) \nonumber \\
 &=& \frac{1}{{2b}}\left\{ {1 - \frac{{1 - b}}{z} - \frac{r(z)}{z}} \right\} . \label{eq:Gsolution}
\end{eqnarray}

\begin{figure}[t]
 \centering
 \includegraphics[height=1.6cm]{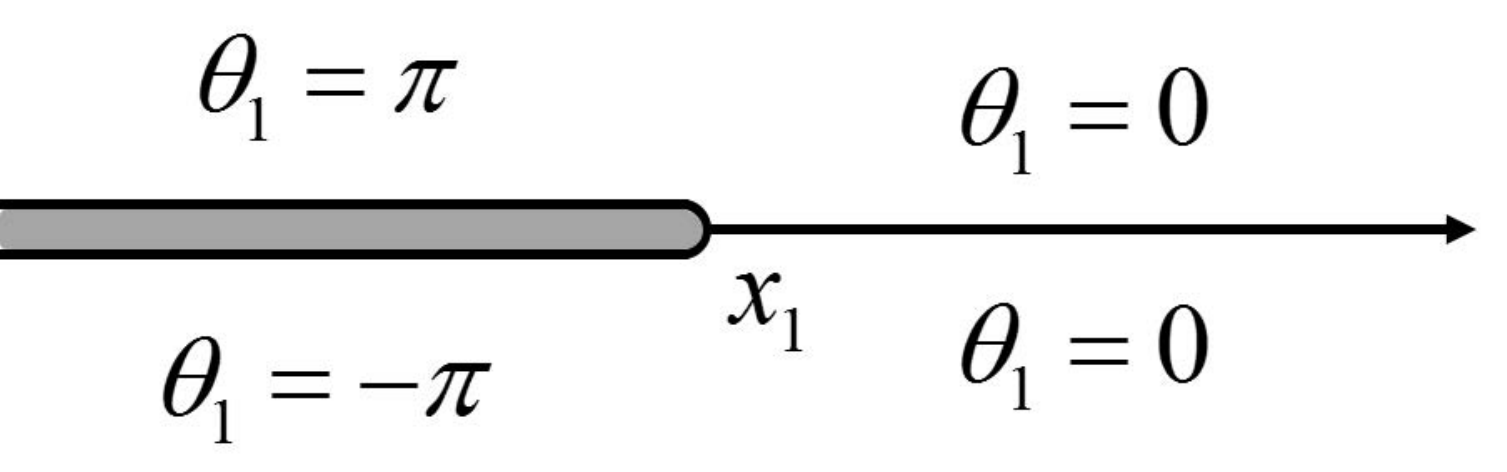}
 \caption{Typical argument value assignment for a single branch point.}\label{fig:singlecut}
\end{figure}

\begin{figure}[t]
 \centering
 \includegraphics[height=2.6cm]{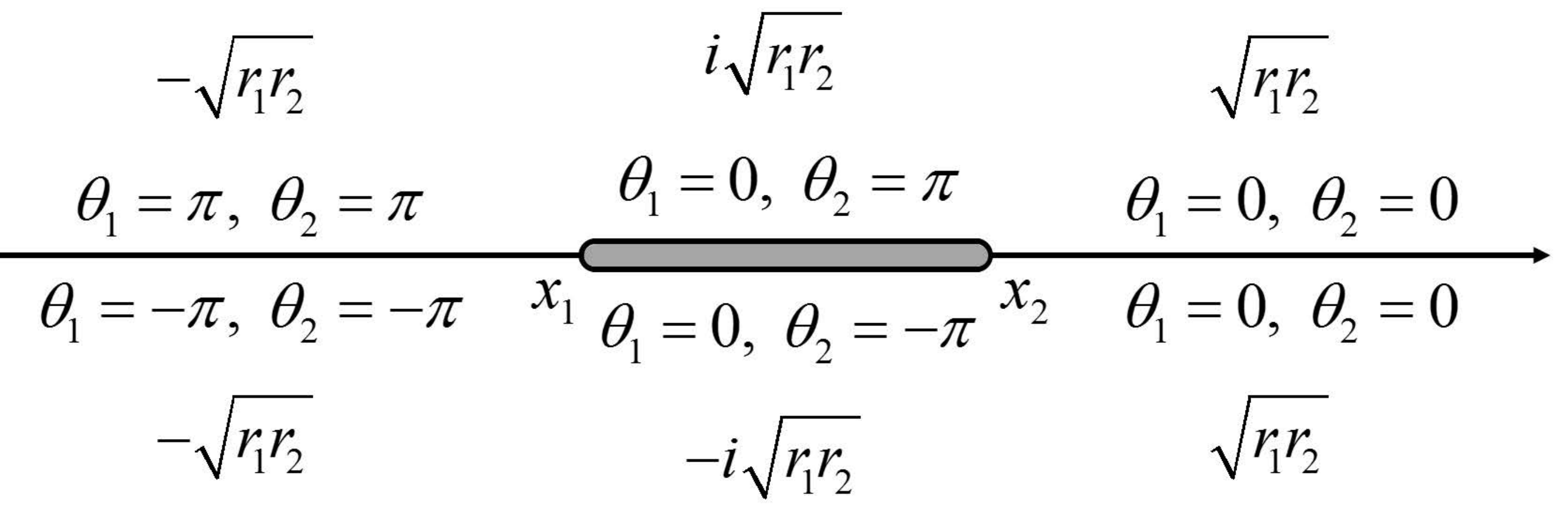}
 \caption{Correct value assignment of arguments $\theta_1$, $\theta_2$ and the resulting value of $r(z)$.}\label{fig:doublecut}
\end{figure}

Recall that ever since Eq.(\ref{eq:expansionB}), we have been working within the vicinity of $z=\infty$. So we need to pick the correct solution to Eq.(\ref{eq:BequationA}) which is the analytically continuation of $B(z)$ from this vicinity. Checking our definition of $G(z)$ (Eq.~\ref{eq:defG}) and $B(z)$ (Eq.~\ref{eq:defB}), we see that both of them should be analytical at $z=\infty$, with the values
\begin{eqnarray}
 \mathop {\lim }\limits_{z\to\infty } G(z) &=& 0 , \\
 \mathop {\lim }\limits_{z\to\infty } {B_1}(z) &=& 1 .
\end{eqnarray}
This requires $\frac{r(z)}{z}$ be analytical at $z=\infty$ with the value
\begin{equation}
\mathop {\lim }\limits_{z\to\infty } \frac{{r(z)}}{z} = 1 . \label{eq:rconfine}
\end{equation}
To find the form of $r(z)$ satisfying these conditions, we first notice that the two solutions to the equation ${\left[ {z - (1 - b)} \right]^2} - 4bz = 0$ are both real positive due to $b>0$. We denote them as
\begin{eqnarray}
 {x_1} &=& {(1 - \sqrt b )^2} , \\
 {x_2} &=& {(1 + \sqrt b )^2} .
\end{eqnarray}
Then
\begin{equation}
 \left[ {r(z)} \right]^2 = {\left[ {z - (1 - b)} \right]^2} - 4bz = (z - {x_1})(z - {x_2}) .
\end{equation}
If we define
\begin{eqnarray}
 z - {x_1} &\equiv& {r_1}{e^{i{\theta _1}}} , \\
 z - {x_2} &\equiv& {r_2}{e^{i{\theta _2}}} ,
\end{eqnarray}
the root $r(z)$ can be written as
\begin{equation}
 r(z) = {\left\{ {{{\left[ {z - (1 - b)} \right]}^2} - 4bz} \right\}^{\frac{1}{2}}} = \sqrt {{r_1}{r_2}} {e^{i\frac{{{\theta _1} + {\theta _2}}}{2}}} .
\end{equation}
Then specifying the branch is just to specify the values of the arguments $\theta_1$, $\theta_2$. For a single branch point, for example $x_1$, a typical assignment of $\theta_1$ would look like Fig.~\ref{fig:singlecut}. But any line starting from $x_1$ ending at $\infty$ can serve as the branch cut. We thus have many choices for each branch cut. However, to make $\frac{r(z)}{z}$ analytical at $z=\infty$, we have to overlap these two branch cuts, both to the left (or equivalently both to the right). The remaining freedom of globally shifting $\theta_1$ or $\theta_2$ by integer multiple of $2\pi$ is fixed by condition Eq.~(\ref{eq:rconfine}). It turns out the correct assignment (Fig.~\ref{fig:doublecut}) is just a repetition of Fig.~\ref{fig:singlecut} applied to both $x_1$ and $x_2$. (If we did not put in a minus sign by hand in Eq.~(\ref{eq:Bsolution}), Eq.~(\ref{eq:rconfine}) would require us to globally shift the assignment of $\theta_2$ (or $\theta_1$) by $\pm 2\pi$ in Fig.~\ref{fig:doublecut}. This would result in the same minus sign for $r(z)$.)

Now we can compute $\rho(E)$. From the solution
\begin{equation}
G(z) = \frac{1}{{2b}}\left\{ {1 - \frac{{1 - b}}{z} - \frac{r(z)}{z}} \right\} ,
\end{equation}
and the branch structure of $r(z)$ (Fig.~\ref{fig:doublecut}), we clearly see that $\mathop {\lim }\limits_{\varepsilon  \to {0^ + }} {\mathop{\rm Im}\nolimits} G(E + i\varepsilon ) = 0$ except when $E$ falls on the branch cut of $r(z)$: $E \in ({x_1},{x_2})$, or $E$ hits the pole of $G(z)$: $E=0$. For the first case, the only contribution to $\mathop {\lim }\limits_{\varepsilon  \to {0^ + }} {\mathop{\rm Im}\nolimits} G(E + i\varepsilon ) = 0$ comes from $r(z)$, and from Fig.~\ref{fig:doublecut} we get
\begin{eqnarray}
 &&\mathop {\lim }\limits_{\varepsilon  \to {0^ + }} {\mathop{\rm Im}\nolimits} G(E + i\varepsilon ) \supset - \frac{1}{{2b}}\frac{{\sqrt {{r_1}{r_2}} }}{E} \nonumber \\
 &=&  - \frac{1}{{2b}}\frac{{\sqrt {({x_2} - E)(E - {x_1})} }}{E}{I_{\{ E \in ({x_1},{x_2})\} }} .
\end{eqnarray}
For the second case, we need to compute the residue of the pole $z=0$
\begin{eqnarray}
 && {\rm{res}}(G(z = 0)) = \frac{1}{{2b}}\left\{ { - (1 - b) - ( - \sqrt {{r_1}{r_2}} )} \right\} \nonumber \\
 &=&  - \frac{1}{{2b}}\left\{ {1 - b - \sqrt {{{(1 - b)}^2}} } \right\} = (1 - \frac{1}{b}) \cdot {I_{\{ b \ge 1\} }} , \nonumber
\end{eqnarray}
which gives
\begin{equation}
\mathop {\lim }\limits_{\varepsilon  \to {0^ + }} {\mathop{\rm Im}\nolimits} G(E + i\varepsilon ) \supset (1 - \frac{1}{b}){I_{\{ b \ge 1\} }} \cdot \left[ { - \pi \delta (E)} \right] .
\end{equation}
Combining the two pieces, we eventually get our result
\begin{eqnarray}
 \rho (E) &=&  - \frac{1}{\pi }\mathop {\lim }\limits_{\varepsilon  \to {0^ + }} {\mathop{\rm Im}\nolimits} G(E + i\varepsilon ) \nonumber \\
 &=& \frac{1}{{2\pi }}\frac{1}{b}\frac{{\sqrt {({x_2} - E)(E - {x_1})} }}{E}{I_{\{ E \in ({x_1},{x_2})\} }} \nonumber \\
 && + (1 - \frac{1}{b})\delta (E){I_{\{ b \ge 1\} }} ,
\end{eqnarray}
with $x_1=(1-\sqrt b)^2$, $x_2=(1+\sqrt b)^2$.

\bibliography{Marchenko_Pastur_Proof_arXiv3}
\bibliographystyle{apsrev4-1}

\end{document}